\documentclass[journal]{vgtc}         

\ifpdf
\pdfoutput=1\relax                   
\usepackage{graphicx}                
\DeclareGraphicsExtensions{.pdf,.png,.jpg,.jpeg} 
\else
\usepackage{graphicx}                
\DeclareGraphicsExtensions{.eps}     
\fi%

\graphicspath{{figures/}{pictures/}{images/}{./}} 

\usepackage{microtype}                 
\PassOptionsToPackage{warn}{textcomp}  
\usepackage{textcomp}                  
\usepackage{mathptmx}                  
\usepackage{times}                     
\usepackage{cite}                      
\usepackage{tabu}                      
\usepackage{booktabs}                  
\usepackage{subfig}
\usepackage{amssymb}
\usepackage{amsmath}
\usepackage{soul,color} 
\usepackage{paralist}
\usepackage{todonotes}
\usepackage{xspace}
\usepackage{todonotes}

\newtheorem{observation}{Observation}

\usepackage{enumitem}
\usepackage{multirow}
\usepackage{booktabs}
\usepackage{tabularx}

\onlineid{0}

\vgtccategory{Research}
\vgtcpapertype{algorithm/technique}

\title{Zoomless Maps: External Labeling Methods for the Interactive Exploration of Dense Point	Sets at a Fixed Map Scale}

\author{Sven Gedicke, Annika Bonerath, Benjamin Niedermann, and Jan-Henrik Haunert}
\authorfooter{
\item Sven Gedicke (gedicke@igg.uni-bonn.de),
\item Annika Bonerath (bonerath@igg.uni-bonn.de),
\item Benjamin Niedermann (niedermann@igg.uni-bonn.de), and 
\item Jan-Henrik Haunert (haunert@igg.uni-bonn.de)\\ 
are with Geoinformation Group, Institute of Geodesy and Geoinformation,
  University of Bonn.}

\newcommand{\attributionStamen}{\textit{Map tiles by \href{stamen.com}{Stamen Design}, under CC BY 3.0. Data by \href{openstreetmap.org}{OpenStreetMap}, under ODbL.}}

\shortauthortitle{Gedicke \MakeLowercase{\textit{et al.}}: Zoomless Maps: External Labeling Methods for the Interactive Exploration of Dense Point	Sets at a Fixed Map Scale}

\abstract{Visualizing spatial data on small-screen devices such as
	smartphones and smartwatches poses new challenges in computational
	cartography. The current interfaces for map exploration require
	their users to zoom in and out frequently. Indeed, zooming and
	panning are tools suitable for choosing the map extent corresponding
	to an area of interest. They are not as suitable, however, for
	resolving the graphical clutter caused by a high feature
	density since zooming in to a large map scale leads to a loss of
	context. Therefore, in this paper, we present new external labeling methods that allow a user to navigate through dense sets of
	points of interest while keeping the current map extent fixed. We provide a unified model, in which labels are
	placed at the boundary of the map and visually associated with the
	corresponding features via connecting lines, which are called
	leaders. Since the screen space is limited, labeling all features
	at the same time is impractical. Therefore, at any time, we label a
	subset of the features. We offer interaction techniques to change
	the current selection of features systematically and, thus, give the
	user access to all features. We distinguish three methods, which
	allow the user either to slide the labels along the bottom side of the map or
	to browse the labels based on pages or stacks.  We present a generic
	algorithmic framework that provides us with the possibility of
	expressing the different variants of interaction techniques as
	optimization problems in a unified way. We propose both exact
	algorithms and fast and simple heuristics that solve the
	optimization problems taking into account different criteria such as
	the ranking of the labels, the total leader length as well as the
	distance between leaders. In experiments on real-world data we
	evaluate these algorithms and discuss the three variants with
	respect to their strengths and weaknesses proving the flexibility of
	the presented algorithmic framework.    	
}

\keywords{external labeling, interactive maps, map exploration, small screens, algorithms, optimization}

\CCScatlist{ 
	\CCScat{K.6.1}{Management of Computing and Information Systems}%
	{Project and People Management}{Life Cycle};
	\CCScat{K.7.m}{The Computing Profession}{Miscellaneous}{Ethics}
}

\teaser{
	\centering
        \includegraphics[page=7,width=0.98\linewidth]{teaser}       
	\caption{An illustration of a smartwatch after querying all
          restaurants in the near surrounding of a user. The three rows of displays show the labeling methods that we present in this paper; \emph{multi-page boundary labeling} topmost, \emph{sliding boundary labeling} centrally and \emph{stacking boundary labeling} bottommost.
         \attributionStamen}
	\label{fig:teaser}
}

\renewcommand{\manuscriptnotetxt}{}
\vgtcinsertpkg
\begin{document}
	\firstsection{Introduction}
	\maketitle
In the last years devices such as smartphones and smartwatches have
conquered our daily life and have made digital maps available at any
time. However, due to the small screen sizes, the presentation of
spatial information on such devices demands the development of new and
innovative visualization
techniques. 
As an example, take a digital map that shows the results of a query for
restaurants in the near surroundings of the user;
see~\autoref{fig:teaser}. Desktop systems and tablets typically offer
enough space to place labels for an appropriately large selection of
restaurants still preserving the legibility of the background map. In
contrast, for small-screen devices---especially for
smartwatches---this is hardly possible as the screen may take
only few labels without covering the map too much.
	
The challenges posed by limited space for label placement are often
softened by interactive map operations such as panning and
zooming. They provide the user with the possibility of exploring the
map by digging into its details. Hence, when there is too much
information on the screen or the user is interested in some details,
he or she can enlarge the map creating more space for the user's area
of interest. In particular, with such interactive operations no
information is lost, as any label can be displayed by zooming in far
enough. However, this might easily become cumbersome. Furthermore, the
loss of the map's context is an additional major drawback, which is
particularly severe on small-screen devices. While the user enlarges
the area of interest, information about the surroundings gets
lost. In the example of searching for restaurants the user might need
to zoom in and out repeatedly to first find the preferred restaurant
and then to locate it on the entire map.
	
\begin{figure}[t]
		\includegraphics[page=1,width=\linewidth]{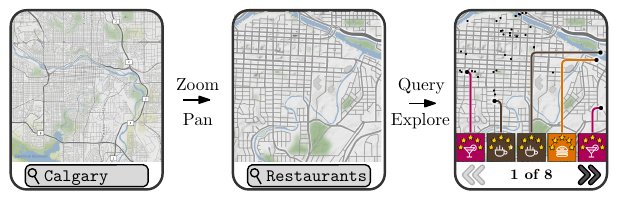}
		\caption{After obtaining
			a broad overview of the city, the user
			zooms in to enlarge the area of interest. Then, the
			user queries for restaurants and explores them without
			further zooming. \attributionStamen }
		\label{fig:zoom}
\end{figure}

In this paper, we investigate labeling methods for \emph{zoomless
  maps}. Such maps reduce the necessity for zooming the map by
providing the user with additional interaction techniques that can be
used for browsing the content of a temporarily fixed section of the
map~\cite{Gedicke2019}. They implement the visual
information-seeking mantra by Shneiderman~\cite{Shneiderman1996}:
\emph{overview first, zoom and filter, then details-on-demand}; see
\autoref{fig:zoom}. In terms of application, starting with an overview
map, the user first zooms and pans the map obtaining an adequate map
of the area of interest (e.g., the city center). In
a second step, the user applies the additional interaction techniques
to browse all information (e.g., all restaurants) contained in the
overview map keeping the view of the map fixed. Hence, the user has an
overview of the entire area of interest at any time without the need
of zooming in for additional information.
	
The labeling methods that we present in this work implement such
additional interaction techniques and allow the user to display all
labels successively without excluding any information beforehand.  In
the running example of searching for restaurants not all labels are
presented at the same time, but the user can consecutively view them
while keeping the displayed map region fixed. Hence, the user can
systematically explore a region and can be finally sure to
have obtained all information.
	
From a visualization point of view, placing the labels internally on the map leads to occluded map content. Especially, when using pictograms or even small images as labels, internal placement can quickly reach its limits in terms of legibility. Moreover, internal labeling does not provide the possibility of presenting labels in a specific order, e.g., with respect to the ratings of the labels' point features.
Hence, we follow the idea to place labels at the bottom of the map in order to avoid hiding map content. We visually associate the labels with their point
features\footnote{As we only consider points as features but no other
  types (e.g., lines and areas), we call point features more shortly \emph{features} in the
  remainder. } by connecting them via a thin curve, also called
\emph{leader}. Placing the labels alongside the boundary of a
rectangular map or figure is known in the literature as \emph{boundary
  labeling}, which is a special case of \emph{external
  labeling}~\cite{Bekos2019}.  For small-screen devices the number~$k$
of labels that can be placed at the bottom of the map is typically
small -- in our experiments we set $k=5$.  Hence, in contrast to
previous work on boundary labeling, which has mostly considered static
labelings~\cite{Bekos2019}, we create and optimize the labelings for
user interaction. We consider the following three
labeling methods that offer the possibility of interactively
exploring the information by browsing through the labels\footnote{A
  demo of all three labeling methods can be tried out on\\
  \url{https://www.geoinfo.uni-bonn.de/interactive-boundary-labeling}}; for an illustration\footnote{The control bar shall illustrate the interaction by the user and can be omitted.} see \autoref{fig:concept}. Each method offers advantages that can be beneficial in various use cases.
	
        



        \newcommand{\figWidth}{0.28}
        \newcommand{\figScale}{1.25}
\begin{figure}[t]
  \centering
  \begin{tabular}{cccc}
   &\textbf{Multi-Page} & \textbf{Sliding} & \textbf{Stacking} \\
    \begin{minipage}{0ex}\rotatebox{90}{\textbf{Las Vegas}}\end{minipage}& \begin{minipage}{\figWidth\linewidth}\includegraphics[page=4,width=\linewidth]{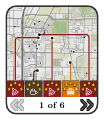}\end{minipage} & \begin{minipage}{\figWidth\linewidth}\includegraphics[page=5,width=\linewidth]{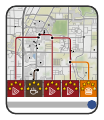}\end{minipage} & \begin{minipage}{\figWidth\linewidth}\includegraphics[page=6,width=\linewidth]{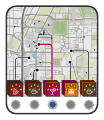}\end{minipage} \\
    \begin{minipage}{0ex}\rotatebox{90}{\textbf{Pittsburgh}}\end{minipage}& \begin{minipage}{\figWidth\linewidth}\includegraphics[page=10,width=\linewidth]{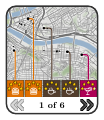}\end{minipage} & \begin{minipage}{\figWidth\linewidth}\includegraphics[page=11,width=\linewidth]{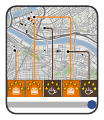}\end{minipage} & \begin{minipage}{\figWidth\linewidth}\includegraphics[page=12,width=\linewidth]{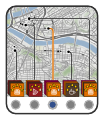}\end{minipage} \\
   & (a) & (b)& (c)\\
  \end{tabular}
  \caption{Illustration of the labeling methods for two instances. (a) The labels are distributed on multiple pages, which the user can browse through. (b) The labels are arranged such that the user can continuously slide the labels from right to
left. (c) The labels are distributed on five stacks. The leaders of
the stack that is currently explored by the user are displayed
simultaneously.  \attributionStamen}
  \label{fig:concept}
\end{figure}

      
\begin{compactenum} 	
\item[L1] \textbf{Multi-page boundary labeling.} This labeling method
  distributes the labels over multiple \emph{pages} such that each
  page consists of a boundary labeling; see \autoref{fig:concept}a. The
  user can navigate through the sequence of pages displaying at most $k$ labels in each
  step. Hence, after
  $\lceil \frac{n}{k} \rceil$ steps the user has obtained all pages
  and labels. Following Gedicke et al.\cite{Gedicke2019}, who
  considered a similar setting for placing labels internally, we call such a
  sequence of pages a \emph{paging} of the labels.  We present an
  algorithm for creating pagings optimizing both the order of the
  labels with respect to the importance of the features as well as the
  leader length. The distribution of labels on pages allows a user to gather  information with a small number of interactions and to quickly locate individual labels.
                
\item[L2] \textbf{Sliding boundary labeling.}  This labeling method
  arranges the labels in a single row that can be continuously slid
  along the bottom of the map; see
  \autoref{fig:concept}b. Only the labels
  directly below the map are displayed to the user.  However, this
  may easily lead to leaders that intersect or closely run in
  parallel. Using a graph-based model, we therefore optimize the order
  of the labels in this sequence taking into account the importance of
  the labels, the number of crossings as well as the vertical distance between
  two~leaders. Since sliding boundary labeling supports continuous animation, users can easily trace changes in the labeling.
  
\item[L3] \textbf{Stacking boundary labeling.} This labeling method
  creates $k$~stacks of labels below the map; see
  \autoref{fig:concept}c. We distribute
  the labels such that each label belongs to exactly one stack and the leader length is minimized. The
  topmost label of the stack is connected to its feature via a
  leader. When the user clicks on the stack the topmost label is
  pushed underneath the bottommost label. Hence, the second topmost
  label moves up and is then connected to its feature. Since each stack can be operated independently, stacking boundary labeling provides the possibility of creating customized labelings.
  
\end{compactenum}
	
From a more technical point of view, we provide a generic framework, which subsumes all three labeling methods. This allows us to define mathematical optimization problems
expressing these labeling methods in a unified way. For all
three variants we present algorithms that yield optimal
solutions. This provides us with the possibility of evaluating the
underlying labeling methods independently from the applied
algorithms. In contrast to previous work on boundary labeling, which
mostly assumes specialized settings (e.g., only $L$-shaped leaders are
allowed, or the algorithms work only for single optimization criteria
such as total leader length), our framework allows us an easy adaptation
to any boundary labeling style and the consideration of multi-criteria
objectives. In our evaluation we point out that none of the labeling
methods prevail the others, but have their pros and cons depending on
the application. We analyze them in greater detail in order
to give the reader a simple way to select an appropriate technique for
the specific purposes. Further, as some of the algorithms are rather
slow and are developed for evaluation purposes only, we also present
simple heuristics that yield near-optimal results fast enough for
interactive operations.
	
The paper is structured as follows. After discussing
related work (Section~\ref{sec:related-work}), we present an expert study
(Section~\ref{sec:expert-study}) on the three labeling methods. We
conducted the study at an early stage of the algorithmic
development to fine-tune the methods and objectives of
the optimization problems. Subsequently, we present a mathematical
model (Section~\ref{sec:model}) that we use to implement our labeling methods (Section~\ref{sec:multi-page}--\ref{sec:stacking}). Finally, we evaluate
our algorithms on real-world data (Section~\ref{sec:evaluation}).
	
\section{Related Work}\label{sec:related-work}


Mobile cartography is a challenging field of research, especially
since the available space and the interaction capacities for
visualizing spatial information are limited.
For presenting spatial content on mobile devices, a broad range
of criteria is considered. In the field of mobile 
applications, Reichenbacher et al.~\cite{reichenbacher2016assessing}
take the geographical context into account and propose methods for
assessing the geographical relevance of spatial objects. Integrating
further context such as physical, temporal and user-dependent
criteria, Pombinho et al.~\cite{pombinho2015adaptive} introduce a
framework for mobile visualization. Usually, the requirements
for the visualization in mobile cartography relate to the
tasks that a user wants to accomplish~\cite{reichenbacher2001adaptive}. For
many specific tasks such as pedestrian
navigation~\cite{perebner2019applying, wenig2015stripemaps} and
annotating map content~\cite{harrie2005algorithm, zhang2006real,
  Gedicke2019} research has already focused on small-screen devices.
Especially in exploration tasks visualizations tend to be visually
cluttered. Korpi and Ahonen-Rainio~\cite{korpi2013clutter} systematically review reduction methods based on cartographic operations
such as selection, aggregation and typification.

Another possibility for clutter reduction is providing the user with interaction techniques.
Zooming is one of the most established and important
interactions in digital maps, as it is both an intuitive and powerful
tool for exploring the content of the map. Thus, the automated
creation of digital maps that provide zooming has been an important
subject of research in computational cartography. New algorithms for
map generalization and label placement have been developed that ensure
certain criteria of consistency over a large range of scales and,
thereby, help users to keep track of the changes that occur in a map
during zooming
\cite{Been2006,Been2010,Barth2016,ChimaniEtAl2014,SesterB04,SubaMO16,OosteromM14,Gemsa2020}.
Further, algorithms for continuously morphing~\cite{NollenburgMWB08}
as well as algorithms for gradually transforming~\cite{Peng2017} two
maps of different scales into each other have been proposed.  Still,
zooming becomes cumbersome when it is the only means for exploring the
details of the map. This particularly drops the last integral part,
namely \emph{details-on-demand}, of the visualization mantra
\emph{overview first, zoom and filter, then details-on-demand} by
Shneiderman~\cite{Shneiderman1996}.
	
Different tools have been proposed for reducing the necessary amount
of zooming.  One of them makes use of a \emph{lens}, e.g., implemented
as a simple circle, which the user can move over the map. Only 
features that are contained in the lens are annotated with labels,
which are placed outside of the lens and are connected by thin leaders
with their features. However, often there is not enough
space in the close surroundings of the lens to show all labels
without overlaps. Fekete et al.~\cite{Fekete1999}, Balata et
al.~\cite{Balata2014} and Heihnsohn et al.~\cite{Heinsohn2014}
therefore place the labels in the extended margins of the lens, which
is suitable for desktop systems but not for
small-screen devices.  Other strategies that keep the labels close to
the boundary of the lens either select a conflict-free subset of
labels~\cite{Fink2012,Haunert2014} or allow the labels to
overlap~\cite{Niedermann2019}.
	
Gedicke et al.\cite{Gedicke2019} take a different strategy and
distribute the labels over multiple pages, placing them on the map
close to their features. This, however, requires rather small
labels to keep the background map legible. We follow the core
idea of showing not all labels at the same time, but enhance this
concept with boundary labeling and interaction. 
	
From an algorithmic point of view, Bekos et al.~\cite{Bekos2004}
introduced the first formal model for boundary labeling, which assumes
that the map is rectangular and the labels are placed alongside of the
boundary of the map. This was the starting point for a plethora of
follow-up works~\cite{Bekos2019}. A vast majority of them present
highly specialized algorithms that are targeted at static figures and
maps without providing any user interaction. As an exception,
N\"ollenburg et al.~\cite{Noellenburg2010} consider boundary labeling
for dynamic maps taking zooming into account. In contrast to previous
work, we present a framework that supports different styles of
boundary labeling without the need to change the underlying
algorithms.  For a detailed survey on external labeling techniques we
refer to~\cite{Bekos2019}.

\newcommand{\methodA}{\textsc{Internal}\xspace}
\newcommand{\methodB}{\textsc{InternalPaging}\xspace}
\newcommand{\methodC}{\textsc{BoundPaging1}\xspace}
\newcommand{\methodD}{\textsc{BoundPaging2}\xspace}
\newcommand{\methodE}{\textsc{BoundStacking}\xspace}
\newcommand{\methodF}{\textsc{BoundSliding}\xspace}

\section{Expert Study}\label{sec:expert-study}
\begin{figure}[tb]
  \centering
  \subfloat[\small \methodA \label{fig:userstudy_1}]
  {\includegraphics[width=0.15\textwidth]{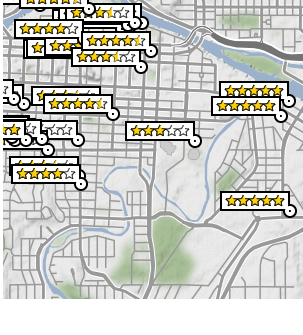}}
  \hspace{5px}
  \subfloat[\small \methodB \label{fig:userstudy_2}]  
  {\includegraphics[width=0.15\textwidth]{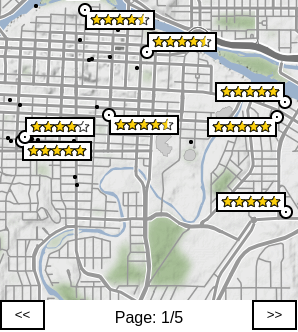}}
  \hspace{5px}
  \subfloat[\small \methodC \label{fig:userstudy_3}]
  {\includegraphics[width=0.15\textwidth]{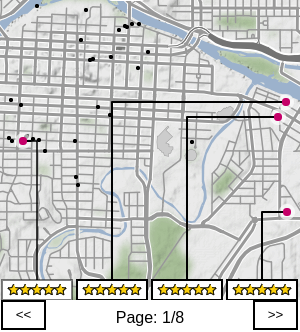}}
  \\
  \subfloat[\small \methodD \label{fig:userstudy_4}]
  {\includegraphics[width=0.15\textwidth]{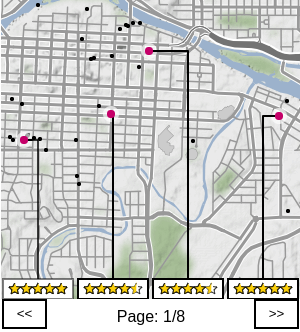}}
    \hspace{5px}
  \subfloat[\small \methodE \label{fig:userstudy_5}]
  {\includegraphics[width=0.15\textwidth]{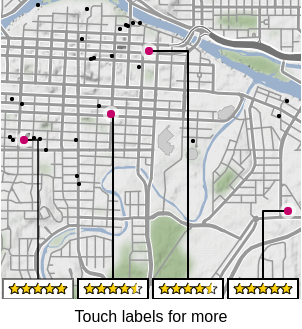}}
    \hspace{5px}
  \subfloat[\small \methodF \label{fig:userstudy_6}]
  {\includegraphics[width=0.15\textwidth]{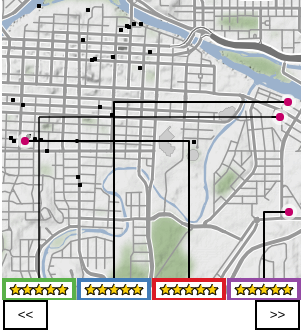}}
		\caption{The methods presented in the expert study. \attributionStamen} 
		\label{fig:userstudy}
\end{figure}            
We conducted a preliminary study on the proposed
visualization techniques.
We aimed at improving our methods and 
verifying whether the design criteria from literature are valid for
the proposed interactions.
By discussing the methods with experts in an early stage, we identified
	inconsistencies before fine tuning
	the algorithms. The study involved nine experts in the
fields of visualization, cartography and computational
geometry. We presented the different visualizations in a
web-browser, simulating a smartwatch\footnote{\url{https://www.geoinfo.uni-bonn.de/interactive-boundary-labeling}}; see \autoref{fig:userstudy}.  We asked the experts to use this
simulation envisioning the following scenario: \textit{A user searches
  for restaurants in Calgary via his or her mobile device. The
  restaurants are labeled with their star rating. In order to maintain
  orientation within the city despite the small screen, all
  restaurants should be labeled without the user having to zoom in.}
We presented six different methods that address this problem. The
first method, which we call~\methodA, simply places all labels with
the bottom-right corner at their features in increasing order
regarding their star rating. 
This method does not resolve conflicts. The second method, which we
call \methodB, is an adaptation of the technique presented by Gedicke et
al.~\cite{Gedicke2019}. This method distributes the labels on multiple
pages and places them internally. We applied a four-position model for the
placement of the labels, i.e., either the upper-left, upper-right,
lower-left or lower-right corner of the label coincides with the feature. For the optimization we took into account both the
distribution of the labels on the different pages and the position of
the label. For optimizing the distribution we maximized the star rating on the first
pages and further maximized the minimal number of labels on a
page~\cite{Gedicke2019}. The third method, which we call \methodC, and the fourth method,
which we call \methodD, use the algorithms for multi-page boundary
labeling presented in this paper. \methodC purely optimizes the star
rating on the first pages, while \methodD optimizes the balance
between star rating and the total leader length.  The fifth method,
which we call \methodE, and the sixth method, which we call \methodF,
present a stacking and  sliding boundary labeling,
respectively. Both were created manually. We emphasize that \methodF
was intersection-free but there were gaps between consecutive labels
to avoid leader crossings.
	
For the simulation we chose a screen size of 300px $\times$ 300px,
which is a common size for smartwatches~\cite{jackson2019smartwatch}. The data set contains 32
restaurants. In contrast to the labels
presented in
Fig.~\ref{fig:teaser}--\ref{fig:concept}, the labels in the study only contained the restaurants' star rating, but no additional symbol. All methods were implemented without  continuous
animations, so that \methodF could not develop its full potential concerning the traceability of changes. To support the user tracing individual labels during interaction, we colored the labels' outlines differently.


We asked the experts to rate the following five statements from 1 (\emph{disagree}) to 4 (\emph{fully agree}).
\begin{compactenum}
\item[S1] \textit{The method gives me quick access to the well rated
  restaurants.}
\item[S2] \textit{The visual association of labels and points is sufficiently clear.}
\item[S3] \textit{The map content needed for orientation, is not too much
  covered.}
\item[S4] \textit{The assignment of the restaurants to front or back pages
  sufficiently reflects the star rating.}
\item[S5] \textit{The shorter guide lines in \methodD lead to an improved
  visualization compared to \methodC.}
\end{compactenum}
We computed the average for each statement; for~S1--S3
see~\autoref{table:expertstudy}, for statement S4 we obtained 3.67 $\pm$ 0.71  for \methodC and 2.72 $\pm$ 0.67  for \methodD, and for S5 we obtained~$2.56 \pm 1.51$.

\begin{table}[tb]
	\centering
	\caption{Averaged values and standard deviations of the nine experts' ratings for the
statements S1--S3.}
    \label{table:expertstudy}
    \begin{tabular}{l c c c c c} 
    	\toprule 
    	\textbf{Experiment}& \textbf{S1} & \textbf{S2}& \textbf{S3}  \\ 
    	\midrule 
    	\methodA & 2.44 $\pm$ 1.24& 2.22 $\pm$ 1.09& 1.44 $\pm$ 0.53 \\ 
    	\methodB & 3.56 $\pm$ 0.88& 3.56 $\pm$ 0.73& 3.11 $\pm$ 0.78 \\ 
    	\methodC & 3.00 $\pm$ 1.22& 2.78 $\pm$ 0.83& 3.83 $\pm$ 0.35 \\
		\methodD & 2.11 $\pm$ 0.60& 3.11 $\pm$ 0.78& 3.94 $\pm$ 0.17 \\ 
		\methodE & 1.78 $\pm$ 0.67& 3.11 $\pm$ 0.93& 3.83 $\pm$ 0.35 \\ 
		\methodF & 2.56 $\pm$ 1.24& 2.78 $\pm$ 0.97& 3.83 $\pm$ 0.35 \\ 
		\bottomrule
	\end{tabular}
\end{table}

Concerning statement~S1 the experts liked \methodB and \methodC most,
and \methodD and \methodE least. This result corresponds to the fact
that for \methodB and \methodC the distribution on different pages
is strictly in the order of the star ratings. For \methodD and \methodE
this is not the case. Again, for~S2 \methodB is best-rated and least favored is
\methodA which allows overlapping labels. For S3 methods using
external labeling performed best, which is most likely explained by the fact that the labels are
placed outside of the map and the leaders merely cover the map.  The
statements S4 and S5 only concern \methodC and \methodD. In the
experts' opinion \methodC fulfills~S4 substantially better than
\methodD. But for \methodD there is still a tendency that S4 is
sufficiently fulfilled. With varying opinions S5 was rated with $2.56$ on average.

Besides rating the statements, we asked the experts to further comment
on the methods. Commonly mentioned was that the
number of labels was rather small and that \methodE was not intuitive.

The study showed that the interactive methods using
	boundary labeling were accepted well by the experts. In comparison to
	\methodB,  the boundary labeling in general performed better
	for~S3, while \methodB performed better for~S1 and~S2. Overall, we
	emphasize that the aim of this study was not to compare
	\methodB against \methodC, \methodD, \methodE and \methodF.
	Both internal and external boundary labeling have
	advantages and disadvantages, but non out-performs the
	other.

As a conclusion we came up with several improvements
for our labeling methods: (1) we visualize the labels as actual
stacks in \methodE; (2) we enrich the information displayed in a label
by additionally visualizing the restaurants' category (the star
rating can be visualized more compact); (3) we use squared
labels; (4) we strongly focus on a strict
order of the star rating; see~\autoref{fig:teaser} and \autoref{fig:concept}. Besides these
general adaptations we have mainly made adaptations to \methodF: (1) we
use a continuous sliding animation, aiming to improve S2; (2) we forbid
gaps and consider intersections as a soft, rather than a hard
constraint which aims to improve S1 knowing that
it might impair S2; (3) we maximize the vertical
distances of neighbored features which aims at improving S2. 
Considering these improvements we obtain the labeling methods L1, L2, and L3.

	
\section{Algorithmic Framework}\label{sec:model}

\begin{figure}
\centering

\tabskip=0pt
\valign{#\cr
  \hbox{%
 \subfloat[general labeling model \label{fig:labeling_model_a}]
  {\includegraphics[width=0.17\textwidth]{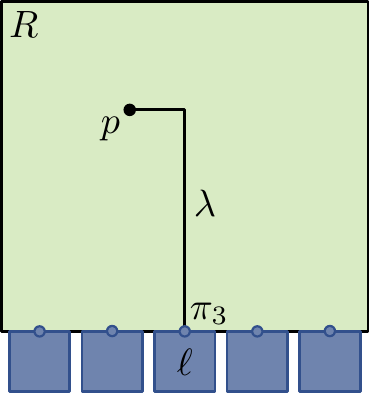}}
  }\cr
  \noalign{\hspace{20px}}
  \hbox{%
\subfloat[multi-page boundary labeling \label{fig:labeling_model_b}]
  {\includegraphics[width=0.23\textwidth]{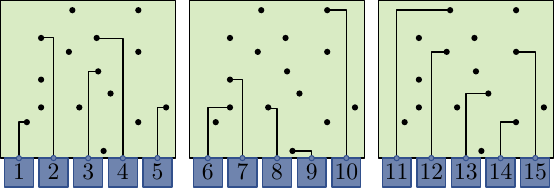}}%
  }\vfill
  \hbox{%
    \subfloat[sliding boundary labeling \label{fig:labeling_model_c}]
  {\includegraphics[width=0.23\textwidth]{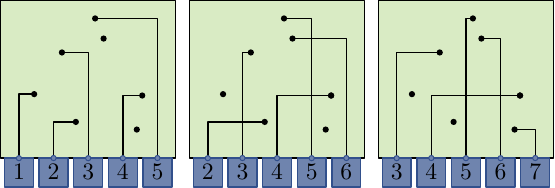}}%
  }\cr
}

\caption{In (a) our general labeling model is shown. The leader that connects feature $p$ with port $\pi_3$ is denoted as $\lambda$. Examples~for multi-page and sliding boundary labeling are shown in (b) and (c), respectively.}

\end{figure}

In this section we introduce an algorithmic framework that serves as a
basis for our three labeling methods L1 -- L3; see Section~1. We
assume that we are given a set $P$ of features within a
pre-defined rectangular region $R\subset\mathbb R^2$ that we refer to
as \textit{map} in the following; see \autoref{fig:labeling_model_a}. We introduce a weight function
$w \colon P \to [0,1]$ that assigns a weight~$w(p)$ to each feature~$p\in P$. Further, each feature in the map is allocated to a
rectangular label $\ell$ that describes the feature. We assume that
all labels are uniformly sized. As the available space on a
small-screen device is limited, we place labels at the bottom side of
the map for $k$ features. More precisely, we define $k$ fixed
positions at the bottom side of the map, which we call
\emph{ports}. We denote the set of ports by
$\Pi = \{\pi_1,\dots,\pi_k\}$ and assume that they are ordered from
left to right. We \emph{attach} a label $\ell$ to a port $\pi_j$ by
placing the label such that the midpoint of its upper side coincides
with $\pi_j$.  The ports are arranged such that the attached labels do
not overlap. Moreover, we visually associate a feature~$p$ and
the port~$\pi$ of an attached label by connecting $p$ with $\pi$ by a
leader. Although the framework covers any kind of leader, we use
so-called \textit{po}-leaders~\cite{Bekos2019} in our experiments. Such leaders consist
of two line segments. The first starts at the feature and is
parallel (p) to the bottom side of the map. The second ends at the
port and is orthogonal (o) to the bottom side of the map.

We model the assignment between features and ports by
a \emph{state} $s\colon\Pi\to P$ that maps each port on a 
feature.  We denote the set of all states by $S$.  Further, a state $s$ \emph{contains} a feature $p\in P$ if a port
$\pi \in \Pi$ exists with $s(\pi)=p$. A state is
\emph{crossing-free} if the leaders that connect the ports with the
contained features of $s$ do not intersect each other.

In the proposed labeling methods we describe a \emph{labeling} as a
sequence $\mathcal S=(s_1,\dots,s_l)$ of $l$ states. We require that
each feature~$p$ is contained in at least one state of
$\mathcal S$. Further, we call a labeling \emph{crossing-free} if all
its states are crossing-free.  Based on the knowledge gained from the
expert study, we consider four
criteria for labelings.
\begin{compactitem}
\item[C1]  Important features should be contained in the first states. 
\item[C2] Crossings of leaders should be avoided.
\item[C3] Vertical distances between the horizontal segments of
  different leaders in the same state should be large.
\item[C4] Leaders should be short.
\end{compactitem}
Depending on the labeling method, we strictly enforce these criteria
or express them as cost functions rating each state $s_i$
with $1\leq i \leq l$ separately. We discuss the cost functions in the following. 

Criterion C1 is described as \emph{weight cost} $c_{\mathrm{W}}(\mathcal S)=\sum_{i=1}^lc_{\mathrm{W}}(s_i)$ with
  \[
   c_{\mathrm{W}}(s_i)=\frac{1}{k \cdot 2^i} \sum_{\pi \in \Pi} \big( 1-w(s_i(\pi)) \big).
 \]
Important features preferably occur in a state with a small index $i$, while less important features are assigned to states with higher indices. 
 
Criterion C2 is described as \emph{crossing cost} $c_{\mathrm{C}}(\mathcal S)=\sum_{i=1}^lc_{\mathrm{C}}(s_i)$ with
\[
  c_{\mathrm{C}}(s_i) = \frac{\text{cross}(s_i)}{{k \choose 2}}. 
\]
The term $\text{cross}(s_i)$ denotes the number of crossings between leaders in
state $s_i$. We normalize the number of crossings by the maximum number
$ {k \choose 2}$ of possible crossings per state.

Criterion C3 is described as \emph{distance cost} $c_{\mathrm{D}}(\mathcal S)=\sum_{i=1}^lc_{\mathrm{D}}(s_i)$ with
\[
  c_{\mathrm{D}}(s_i) = \frac{1}{{k \choose 2}} \sum_{\{p,q\}\in H(s_i)}\frac{1}{|y(p) - y(q)|}.
\]
The set $H(s_i)$ contains all pairs $\{p,q\}$ of features whose
leaders in $s_i$ have horizontal segments partially running above each other. Hence, states with
horizontal leader segments running close above each other have higher distance costs than states
with well separated leaders. We average the cost by the maximal
size ${k \choose 2}$ of $H(s_i)$.

Criterion C4 is described as \emph{leader cost} $c_{\mathrm{L}}(\mathcal S)=\sum_{i=1}^lc_{\mathrm{L}}(s_i)$ with
\[
  c_{\mathrm{L}}(s_i)=\frac{1}{k \cdot 2^i}  \sum_{\pi \in \Pi} \frac{\textrm{leader-length}(\pi,s_i(\pi))}{\text{screen width + screen height}}.
\]

The term $\textrm{leader-length}(\pi,s(\pi))$ denotes the length of
the leader that connects the port $\pi$ with the feature
$s_i(\pi)$. We normalize the length of the leaders by the screen width and height. The relevance of a state decreases exponentially with increasing index $i$, so that short leader lengths are given more priority on front states than on back states.

For each of the three labeling methods we use $c_{\mathrm{W}}$,
$c_{\mathrm{C}}$, $c_{\mathrm{D}}$ and $c_{\mathrm{L}}$ to compose a
bicriteria cost function $c$ that balances two of those cost functions by means of a factor $\alpha\in [0,1]$. We call $c(\mathcal S)$ the \emph{cost} of a labeling $\mathcal S$.
We restricted ourselves to bicriteria cost functions to be capable of controlling the effects of the optimization on the resulting labeling. More precisely, we use the two criteria that affect each method most as soft constraints and express the others as hard constraints if this is reasonable. We argue for the actual choice in the following sections.

\section{Multi-page boundary labeling}\label{sec:multi-page}
In multi-page boundary labeling  the labels are distributed on multiple pages
such that each page labels a distinct set of features (see
\autoref{fig:labeling_model_b}).  Each page corresponds to a state in
a labeling $\mathcal S=(s_1,\dots,s_l)$ with $l = \lceil n/k \rceil$;
in the case that $n/ k \notin \mathbb{N}$, we introduce dummy features
that do not influence the overall solution. We say~$\mathcal S$ is a
\emph{multi-page boundary labeling} if each feature is contained in
exactly one state and it is crossing-free (C2). Among all
multi-page boundary labelings, we search for one that optimizes the
weight cost (C1) and the leader cost (C4). We do not consider the
vertical distances between features (C3), as we deem such distances
mostly problematic in combination with crossings. Based on
$c_{\mathrm{W}}$ and $c_{\mathrm{L}}$, we define the cost of a
multi-page boundary labeling~$\mathcal S$ as
\[
  c_{\mathrm{MPL}}(\mathcal S,\alpha) = \alpha \cdot c_{\mathrm L}(\mathcal S) + (1-\alpha)\cdot c_{\mathrm W}(\mathcal S),
\]
where $\alpha \in [0,1]$ balances the leader and weight cost of $\mathcal S$.
For $\alpha=0$ and $\alpha=1$, only the feature cost and the leader
cost are considered, respectively.  For a single state $s_i$ of
$\mathcal S$ we further define
$c_{\mathrm{MPL}}(s_i,\alpha)=\alpha\cdot c_{\mathrm L}(s_i) +
(1-\alpha)\cdot c_{\mathrm W}(s_i)$. Due to the linearity, we have
$c_{\mathrm{MPL}}(\mathcal S,\alpha)=\sum_{i=1}^l
c_{\mathrm{MPL}}(s_i,\alpha)$.

We show that the problem of finding a multi-page boundary labeling can
be transformed into finding a perfect matching in a bipartite
graph. At first, we observe that for a set of $k$ features and $k$
ports the problem reduces to computing a one-sided boundary labeling
with minimum total leader length for a single page. Such a labeling
can be computed in $O(k \log k)$ time~\cite{benkert2009algorithms} --
no matter whether the length of the leader is normalized or not.  We
reformulate this as follows.

\begin{observation}\label{obs:mpl}
  Let $p_1,\hdots, p_k$ be features and $\pi_1,\hdots, \pi_k$
  label ports. Let $s$ be a not necessarily crossing-free state with
  minimal costs $c_{\mathrm{MPL}}(s,\alpha)$ over all states of
  $p_1,\hdots, p_k$ and $\pi_1,\hdots, \pi_k$. There exists a
  crossing-free state $s'$ with
  $c_{\mathrm{MPL}}(s,\alpha) = c_{\mathrm{MPL}}(s',\alpha)$.
\end{observation}

Observation~\ref{obs:mpl} holds because the number of
features corresponds to the number of ports. Hence, only the leader cost
is considered, but not the weight cost.  Using Observation~\ref{obs:mpl}, we divide
the computation into two steps; first we find
a cost-minimal possibly non crossing-free solution, and in the second step we resolve the crossings on each page.

The first step corresponds to finding a perfect
matching in a bipartite graph. In general, let $G=(V,E)$
be a weighted bipartite graph, i.e.,  $V$ is partitioned into two
disjoint sets $V_1$ and $V_2$ and each edge $e \in E$ connects a
vertex $v_1 \in V_1$ with a vertex $v_2 \in V_2$. A \emph{perfect
  matching} in $G$ is a subset of $E$ such that each vertex is
incident to exactly one edge of the subset. A \emph{minimum-weight} perfect
matching is a perfect matching that minimizes 
the sum of the weights over all selected~edges.

\begin{figure}[tb]
		\centering
		\includegraphics[page=5]{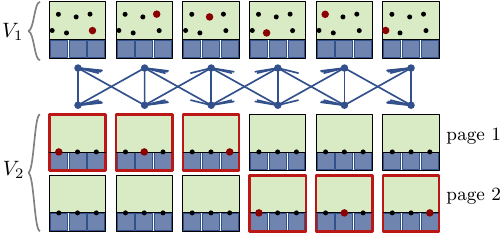}
		\caption{Multi-page boundary labeling visualized as a perfect
                  matching.}
		\label{fig:bipartitperfectmatching}
\end{figure}

For multi-page boundary labeling
the vertex set $V_1$ consists of the vertices $v_p$ where $p$ is a
feature and the set $V_2$ consists of the vertices $v_{\pi,s}$
where $\pi,s$ is a pair of a port $\pi$ and a state $s$; 
see~\autoref{fig:bipartitperfectmatching}. Further, $G$ is
a complete bipartite graph. For each edge $e$ that connects the
vertices $v_p$ and $v_{\pi,s}$, we set the weight equal to 
\begin{align*}
	\frac{1}{2^i}\cdot (1-w(p)) 
	+ \alpha \cdot \frac{1}{2^i} \cdot \frac{\textrm{leader-length}(\pi,s_i(\pi))}{\text{screen width + screen height}}
\end{align*}	
To solve the multi-page boundary labeling allowing crossings, we compute a perfect minimum-weight matching in $G$, which can be done in $O(n^3)$ time
\cite{kuhn1955hungarian,munkres1957algorithms}. In our
implementation, we used a linear programming approach, which is also a common method
\cite{grotschel1985solving}.

In the second step, we apply the algorithmic approach presented by
Benkert et al.  (2009)\cite{benkert2009algorithms} for each state to
find a crossing-free labeling with minimal leader length. We also use
this approach for the stacking boundary labeling and discuss it in
Section~\ref{sec:stacking} in greater detail.

%

\section{Sliding boundary labeling}\label{sec:sliding}
In sliding boundary labeling the labels can be moved along the bottom side of the map.  From a
technical point of view, we assume that the labeling is given as a
sequence $(s_1,\dots,s_l)$ of states; see \autoref{fig:labeling_model_c}. The interaction of sliding the
labels then corresponds to an animated transition between consecutive
states. To guarantee a continuous flow of
labels, we require that from one state $s_i$ to the next state
$s_{i+1}$ the labels are shifted to the left by one port and a new
label appears at the rightmost port. More formally, we say that the
transition from $s_i$ to $s_{i+1}$ is \emph{valid}, if
$s_i(\pi_j) = s_{i+1}(\pi_{j-1})$ for all $1 < j \leq k$ and
$s_i(\pi_1) \neq s_{i+1}(\pi_k)$.




Since---except for the last $k-1$ labels---each label is shifted once from the rightmost to the leftmost port, we do not consider the leader length (C4). Instead, we aim for a labeling that minimizes both the crossing cost (C2) and the distance cost (C3). For $\alpha\in[0,1]$ we define the cost of a labeling~$\mathcal S$ as 
\begin{align*}
c_{\mathrm{Slid}}(\mathcal S) = \alpha \cdot c_{\mathrm{C}}(\mathcal S) + (1-\alpha) \cdot c_{\mathrm{D}}(\mathcal S).
\end{align*}
In the following, we present an exact approach that yields the optimal solution as well as a fast and simple
heuristic.

\subsection{Exact Approach} \label{sec:sliding:ilp}

For obtaining an optimal sliding boundary labeling we
use a graph-based method by modeling our optimization problem as an
constrained \textit{orienteering problem}
\cite{vansteenwegen2011orienteering}. Given a graph in which each
vertex has a score and each edge a length, a source and a target
vertex, the orienteering problem asks for a path from the source to
the target such that the total score along the path is maximized and a
given length is not exceeded. With some adaptations on the original
problem definition this path represents a labeling.
We define an integer linear programming (ILP) formulation that
expresses the problem of finding an optimal labeling as a linear
objective function subject to a set of linear constraints.  In general,
solving such formulations is NP-hard \cite{Garey1979}. However, there exist solvers
that handle many of such formulations in adequate time. We
present a formulation that models the problem as a \textit{flow
  network}. The idea is to send a unit from a source
to a target vertex via the graph.

\begin{figure}[tb]
		\centering
		\includegraphics[page=2,width=0.47\textwidth]{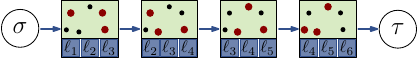}
		\caption{Illustration of an exemplary path from $\sigma$ to $\tau$ in the graph $G$.}
		\label{fig:sliding_graph}
\end{figure}

 
We present now the details of our approach. Let
$G=(S\cup\{\sigma,\tau\},E)$ be a directed graph containing a vertex
for each state $s \in S$. Further, $G$ contains a source vertex
$\sigma$ and a target vertex $\tau$.  Each state $s \in S$ is connected
with the source by an edge $(\sigma,s)$ and with the target by an edge
$(s,\tau)$. Moreover, a pair of states $s,t \in S$ is connected by a
directed edge $(s,t) \in E$ if and only if the transition from state
$s$ to state $t$ is valid. We aim for a path from $\sigma$ to $\tau$ representing
a labeling; see~\autoref{fig:sliding_graph}.

Next, we describe our ILP formulation. For each directed edge
$(s,t)\in E$ we introduce a binary variable $x_{s,t} \in \{0,1\}$. We
interpret $x_{s,t}$ such that $x_{s,t}=1$ if and only if the
transition of state $s$ to state $t$ is used in the
labeling. Additionally, we introduce a variable
$y_{s}\in \{0,\dots,n\}$ for each state $s$ with $n$ being the number
of states. We use $y_{s}$ to ensure that the sub-graph defined by the
edges $e\in E$ with $x_{e}=1$ is actually one connected component
forming a path from $\sigma$ to $\tau$.  We denote the set of all
states that contain a feature $p$ by $S_p$. Further, we define
$E_p\subseteq E$ to be the set that contains any edge $e=(s,t)$ whose
source~$s$ does not contain $p$, but whose target $t$ contains $p$,
i.e., $E_p= E \cap \left((S \setminus S_{p}) \times S_{p} \right)$.
In order to model a path in $G$ that represents a labeling we
introduce the following constraints.
\noindent\begin{minipage}{.5\linewidth}
  \begin{align}
    \sum_{t\in S} x_{\sigma, t}=1 \label{eq:constraint1}\\
    \sum_{t \in S \atop t \neq s} x_{s, t} \leq 1 \ \forall s \in S  \label{eq:constraint3}
  \end{align}
\end{minipage}%
\begin{minipage}{.5\linewidth}
  \begin{align}
   \sum_{s \in S} x_{s, \tau}=1  \label{eq:constraint2}\\
    \sum_{s \in S \atop s \neq t} x_{s, t}=\sum_{u\in S \atop u \neq t} x_{t, u}\  \forall t \in S \label{eq:constraint4}    
  \end{align}
\end{minipage}
\begin{align}
y_{s}+1 \leq y_{t}+(n-1) \cdot\left(1-x_{s, t}\right) & \ \forall s,t \in S, s \neq t \label{eq:constraint5} \\
\sum_{(s,t) \in E_p} x_{s, t}=1 & \ \forall p \in P  \label{eq:constraint6}
\end{align}
Constraint~\ref{eq:constraint1} and Constraint~\ref{eq:constraint2}
ensure that the path begins at the source~$\sigma$ and ends at the
target~$\tau$ by enforcing that exactly one transition from $\sigma$
and one transition to $\tau$ is used.  Hence, in terms of the flow
network, exactly one unit leaves the source and one unit reaches the
target.  Constraint~\ref{eq:constraint3} ensures that each state can
be the origin of at most one transition, i.e., at most one unit leaves
each state. To preserve the flow, Constraint~\ref{eq:constraint4}
implies that if a unit reaches one state, it also leaves this
state. As we aim for one connected path through $G$, we introduce
Constraint~\ref{eq:constraint5}, which guarantees that the states are
numbered in ascending order along the path. Hence,  no cycles can be created. To ensure that the path
corresponds to a labeling, Constraint~\ref{eq:constraint6} guarantees
that each feature $p \in P$ is labeled at least in one state $s$ and
that it only appears in consecutive states.  Subject to
Constraint~\ref{eq:constraint1}--\ref{eq:constraint6} we minimize
\begin{align}
\Big( \alpha \cdot \sum_{s \in S} c_{\mathrm{C}}(s) + (1-\alpha) \sum_{s \in S} c_{\mathrm{D}}(s) \Big) \cdot \sum_{t \in S \atop t \neq s}  x_{s,t}. 
\end{align}
Let $Q=\{e \in E \mid x_e=1\}$ be the path from $\sigma$ to
$\tau$. Going along $Q$ we obtain a sequence
$\mathcal S=(s_1,\dots,s_l)$ of states which forms a sliding labeling
such that $c_{\mathrm{Slid}}(\mathcal S)$ is minimal.


\subsection{Heuristic Approach} \label{sec:sliding:heu}

\begin{figure}[t]
  \centering
  \includegraphics[width=0.47\textwidth]{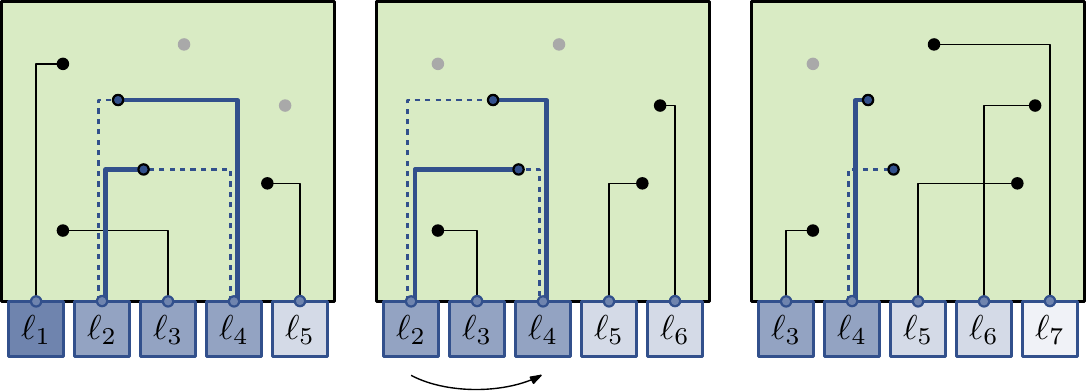}
  \caption{Illustration of a swap. The labels' colors represent the weight of the features. The leaders affected by the swap are shown in blue. The dashed blue leaders correspond to the solution before and the solid blue leaders belong to the solution after the swap. }
  \label{fig:swap}
\end{figure}
	
We present a simple and fast heuristic for sliding boundary labeling
that is based on the local search strategy \textit{hill climbing}.
Starting from an initial solution, such methods iteratively perform
local changes in the current solution to obtain \textit{neighboring}
solutions. Such a neighboring solution is accepted as the new current
solution if it improves the objective. The algorithm terminates if a
pre-defined number of iterations is reached or no more neighboring
solutions of higher quality~exist.

For the initial solution we apply a  simple first-fit strategy:
we create an arbitrary order $p_1,\dots,p_n$ of the features and
define a state $s_i$ as $s_i(\pi_1)=p_i,\dots,s_i(\pi_k)=p_{i+k}$ for
$1\leq i \leq n-k$. Hence, sliding the labels from right to left, the
features are labeled in that order.
%
We iteratively improve that labeling by
performing local changes. We obtain a neighboring
solution by \textit{swapping} the assignments $s_i(\pi) = p$ and
$s_j(\pi') = q$ of two randomly chosen features $p$ and $q$,
i.e., after this \textit{swap} we have $s_i(\pi) = q$ and $s_j(\pi') = p$. For
an illustration of a swap see \autoref{fig:swap}.  If
such a swap improves the value of the objective function, it is
applied obtaining a new solution. The algorithm continues with
performing such swaps until a pre-defined number of iterations is
reached or no more neighboring solutions of higher quality exist.

	\section{Stacking boundary labeling}\label{sec:stacking}	
	The stacking boundary labeling is characterized by its
        interaction technique. Instead of having one pre-defined
        sequence of states, the features are partitioned into $k$
        groups; each assigned to one stack. Hence, we have an own sequence of states for
        each single stack, where each state
        consists of a single feature.
	By clicking on one label, its port is connected to
	the next feature in the stack.
		
	We call $t_\pi \colon i \to P$ with $1 \leq i \leq l$ the
        \emph{stack} of port $\pi$. A feature $p$ is \emph{contained}
        in a stack $t$ at position $i$ if $t(i)= p$. A \emph{stacking
          boundary labeling} consists of the set of stacks
        $\mathcal T=\{t_{\pi_1},\hdots,t_{\pi_k}\}$. We require 
        that the features are evenly distributed over all
        stacks\footnote{We assume $n/k \in \mathbb{N}$. If this is not
          the case we insert dummy feature points.}, i.e., $l = n/k$,
        that each feature is contained in exactly one stack at one
        position, and that the labeling is crossing-free
        (C2). Among all stacking boundary labelings we search for the
        one that optimizes the weights of the features on the first
        stack positions (C1) and the length of the leaders (C4). As there exists a state for every two features of different stacks in which they are labeled simultaneously, we do not consider the vertical distance (C3).  The user can improve
        the visualization by clicking on the labels of features that are vertically close together. 

\begin{figure}[tb]
		\centering
		\subfloat[\label{fig:stacking-labelingA}]
		{\includegraphics[width=0.17\textwidth,page=3]{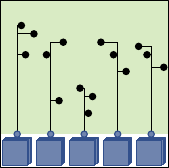}}
		\hspace{20px}
		\subfloat[ \label{fig:stacking-labelingB}]
		{\includegraphics[width=0.17\textwidth,page=4]{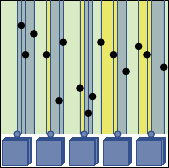}}
		\caption{(a) All leaders of the stacking boundary labeling. (b) Map partitioned into strips (yellow strips: rightwards, blue strips: leftwards)~\cite{benkert2009algorithms}.} 		
		\label{fig:stacking-labeling}
	\end{figure}

	As the primary objective for partitioning the features
        into $k$ groups, we optimize the leader length globally for
        $\mathcal T$ as follows.
	\begin{equation*} 
	\min\ c(\mathcal T) = \min \sum_{\pi \in \Pi} \sum_{i = 1}^l \textrm{leader-length}(\pi,t_\pi(i)) \label{eq:stacking-opt-1}
	\end{equation*}
	As the secondary objective for each sequence, we optimize
        Criterion~C1 by ordering the features of each stack by their
        weight.        
	We show that the problem can be transformed into a static
        boundary labeling problem. We introduce each port of our
        stacking boundary labeling problem $l$ times; see
        \autoref{fig:stacking-labelingA}. Further, we adapt the
        definition of a crossing; we say that two leaders that start
        at the same port cannot cross. A solution of the first step of
        the stacking boundary labeling problem is a crossing-free
        assignment with minimal leader length between the labels' ports
        and features. The approach by Benkert et
        al.~\cite{benkert2009algorithms} solves this problem in
        $O(n \log n)$ time. In the approach the map is partitioned into strips induced by vertical lines
        through each label's port and feature; see
        \autoref{fig:stacking-labelingB}. The strips are categorized
        into leftwards and rightwards directed strips by a comparison
        of the number of labels and the number of features to
        the right and left. Each consecutive set of leftwards
        directed strips is solved with a sweep line approach from right to
        left. Each time the sweep line passes a feature,
        it is inserted into a list $W$ which is ordered by the features'
        $y$-coordinates. Every time the sweep line
        intersects a port, this port is assigned to the
        lowest feature contained in $W$. The rightwards directed
        strips are solved with a symmetric approach. Finally, we sort
        each stack $t_\pi$ by the weight of the features.
	
	A stacking labeling corresponds to a
	multi-page labeling by assigning the $i$-th label of each stack to the $i$-th page, i.e., $s_i(\pi) = t_\pi(i)$.
	
	\section{Evaluation}\label{sec:evaluation}
	Following, we present experiments evaluating our labeling
        methods on real-world data focusing on quantitative
          criteria and performance.

        We
        simulated the use case of querying restaurants with a
        smartwatch visualizing the result on a map.
        At first, we evaluate each labeling method individually. In
        particular, we identify a suitable choice of $\alpha$
        describing the sweet spot between the contradictory optimization
        criteria.  Further, we show that it pays off to optimize the
        criteria C1--C4 by comparing the algorithms' results with the
        worst and best cases considering each criterion. For a cost function~$c$ and two labelings $\mathcal S$ and $\mathcal S'$ we
        define their \emph{relative cost} as
        \begin{equation*}
          \delta(c,\mathcal S,\mathcal S') = \frac{c(\mathcal S)- 
            c(\mathcal S')}
          {c(\mathcal S')}\cdot 100\%.
        \end{equation*}	
        We use the relative cost to compare a labeling $\mathcal S$
        with another labeling~$\mathcal S'$ that is optimal with
        respect to one or multiple of the criteria C1--C4.  Finally,
        we compare the labeling methods with each other emphasizing
        their strengths and weaknesses.

        We use the following experimental setup.  For the smartwatch
        we assume a screen size of 300px~$\times$~300px, which is a
        common resolution~\cite{jackson2019smartwatch}. The
        displayed  map comprises roughly
        5733m~$\times$~5733m. This extent is a good compromise
        between a broad overview of the area of interest and the
        possibility of identifying the location of the restaurant.

        We obtained the data of the restaurants from
        Yelp\footnote{www.yelp.com} for the cities Calgary~(CA), Las
        Vegas~(US), Montreal~(CA), Pittsburgh~(US) and
        Toronto~(CA). Each restaurant is given with its location and
        star rating. The rating varies between one and five;
        half-stars are also allowed. We normalize the ratings
        obtaining weights in the interval $[0,1]$. We create 100
        instances of different map sections each having 30 randomly sampled features; see
        \autoref{fig:concept} for examples.  We use five labels
        ($k=5$), each with a size of 60px$\times$60px. To
        make the visualization more realistic we extend the
        data with categories; e.g., pizza
        restaurant. We note that this has no impact on the algorithms
        nor the evaluation.

The implementations were done in Java, and the ILP and
LP formulations were solved by
Gurobi\footnote{www.gurobi.com}~8.1.0. We ran the experiments
on a Intel(R) Xeon(R) W-2125 CPU clocked at 4.00GHz with 128 GiB
RAM. Considering a server-client communication as the use case for our algorithms, we performed the computations on a server system.



        

\subsection{Multi-page Boundary Labeling}\label{sec:eval_mpl}	
For the multi-page boundary labeling, we evaluate the weight cost (C1), the
leader cost (C4), and the distance cost (C3) for
different values of~$\alpha$. The weight cost and 
leader cost are considered in the objective where~$\alpha$ balances those. 
For the experiments, we
sampled~$\alpha$ in the range~$[0,1]$ with a step
width of~$0.025$. We denote
the set of the resulting~$41$ values by~$\mathcal A$ and a 
multi-page boundary labeling for $\alpha$ by~$\mathcal
S_\alpha$.

	
\begin{figure}[tb]
	\centering
	\includegraphics[width=0.45\textwidth]{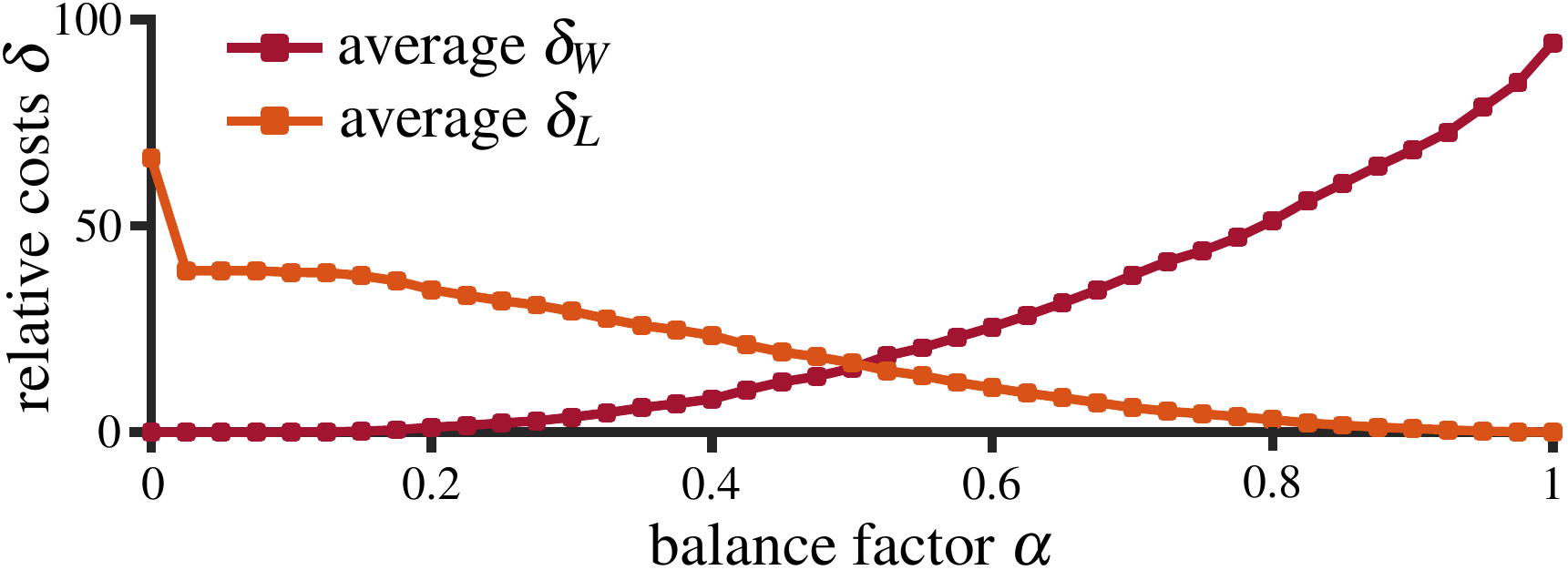}
	\caption{Average relative leader cost $\delta_{\mathrm{W}}$ and average relative feature cost $\delta_{\mathrm{L}}$ for each $\alpha \in \mathcal A$.}
		\label{fig:multi-page-scoreLeader}
\end{figure}

\paragraph{Optimized Criteria C1 and C4}
We consider the relative weight cost~$\delta_{\mathrm{W}}(\mathcal S_\alpha):= \delta(c_{\mathrm W}, \mathcal S_\alpha, \mathcal S_0)$ of a labeling~$\mathcal S_\alpha$ and the labeling~$\mathcal S_0$ with minimal weight cost; see
\autoref{fig:multi-page-scoreLeader}.
Similarly, we define the relative leader
cost~$\delta_{\mathrm{L}}(\mathcal
S_\alpha):=\delta(c_{\mathrm{L}},\mathcal S_\alpha, \mathcal
S_1)$ for a labeling $\mathcal S_\alpha$ and the
labeling~$\mathcal S_1$ being the solution with minimal
leader cost~$c_{\mathrm{L}}$. 
We obtain that the difference between the relative leader
cost for~$\alpha=0$ and~$\alpha= 0.025$ is remarkably large.
When comparing this to the average~$\delta_{\mathrm{W}}$ it
is noticeable that~$\delta_{\mathrm{W}}$ is still
optimal for $\alpha=0.025$. Hence, by
choosing~$\alpha=0.025$ instead of $\alpha=0$, the cost for
$c_{\mathrm W}$ is still minimal, while $c_{\mathrm L}$ is
also considered. Thus, when ordering the features
by their weights is strictly required---as
derived from the expert study---we recommend to
choose~$\alpha=0.025$. Moreover, the relative costs of both criteria intersect at $\alpha=0.5$.
Hence, we deem~$\alpha = 0.5$ to be a suitable compromise
between optimizing the weight cost and leader cost.

\paragraph{Non-optimized Criterion C3}
For the multi-page boundary labeling, we do not consider the distance
cost in the objective. We analyze all pairs $H$ of leaders which are
labeled on the same page and for which the horizontal parts of the
leaders run above each other. At first, we evaluate the size
of $H$ in relation to the overall number of leader pairs on the same
page. On average $20.7\%$ of the leader pairs on the same page
horizontally run above each other. In particular, the ratio
is similar for all values of $\alpha$. Moreover, we analyze the ratio
of leader pairs in $H$ with a vertical distance smaller than $5$ pixels
with respect to the size of $H$.  This ratio varies between $11.0\%$
for $\alpha=0$, and $38.4\%$ for $\alpha=1$. This means, that at most
$38.4\%$ of the $20.7\%$ of leader pairs which have horizontal
segments lying above each other, have a vertical distance
under $5$px. Since the multi-page boundary labeling provides static
pages, this result seems to be appropriate to distinguish the leaders.
 

\subsection{Sliding Boundary Labeling}
Our study (see Section~\ref{sec:expert-study}) showed that experts generally recommend to strictly label the features in descending weight order.
Hence, we consider a special case of sliding boundary labeling in which we enforce Criterion~C1 as hard constraint. While the overall weight order is preserved, we optimize the order within groups of features having the same weight regarding the cost function~$c_{\mathrm{Slid}}(\mathcal S)$.
We modify our exact approach such that the graph $G=(S',E)$ only contains states satisfying Criterion~C1. Hence,  $S' \subseteq S$ is the set of states in which the features are assigned to the ports from left to right in descending weight order.

We also adapted the heuristic as follows. When constructing an initial
solution, we sort the given features in descending weight order. Further,
when obtaining neighboring solutions, we only swap the assignments
$s_i(\pi) = p$ and $s_j(\pi) = q$ of two randomly chosen
features $p$ and $q$ if $w(p) = w(q)$. We use $5000$ iterations as
termination criterion, which has proven to be a suitable choice in
preliminary experiments. For each instance and each considered
weighting $\alpha$ we run our heuristic five times and build the
arithmetic mean of the results.

\begin{figure}[tb] \centering
\includegraphics[width=0.48\textwidth]{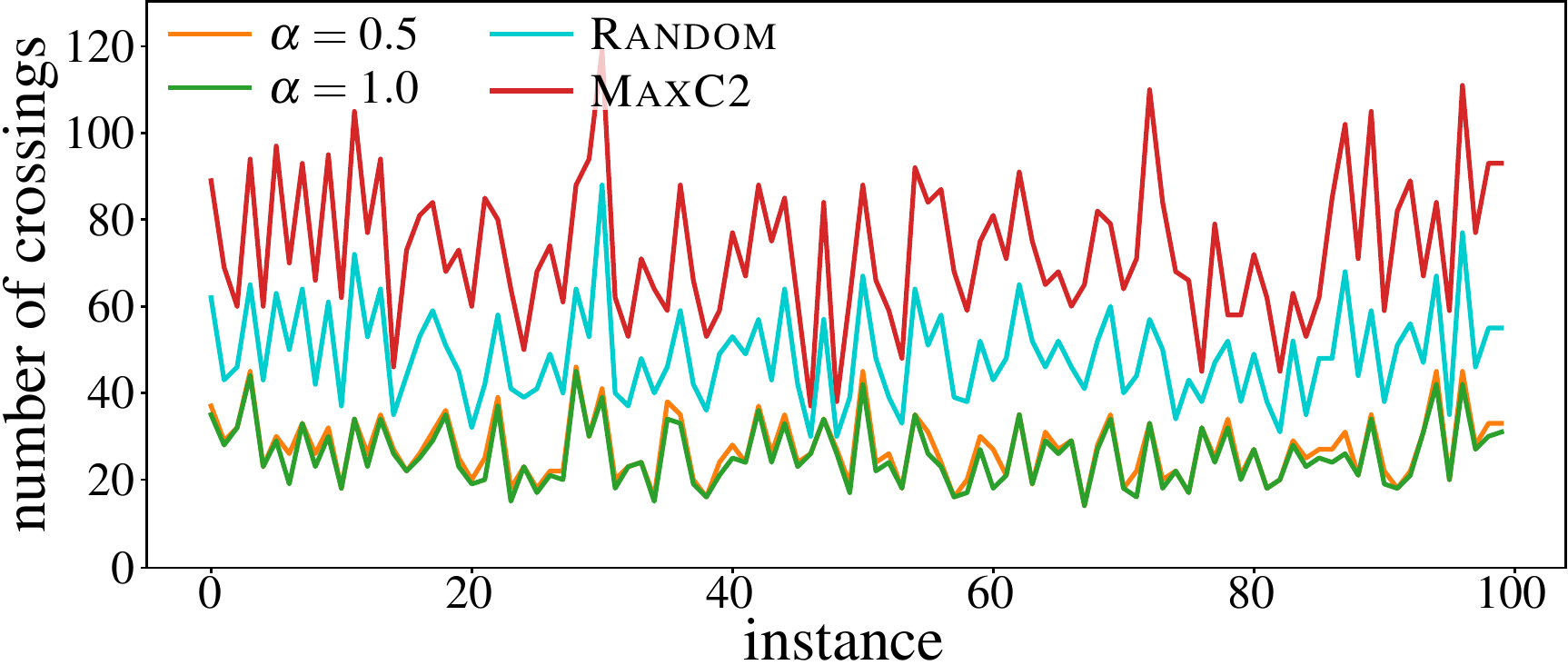}
\caption{Optimal crossing cost $c_{\mathrm{C}}$ (without
  normalization, i.e., number of crossings) for each instance and $\alpha \in \{0.5, 1
  \}$. \textsc{Random} is the number of crossings of randomly generated labelings and \textsc{MaxC2} is the maximal possible number of
  crossings.}
		\label{fig:ilp_crossings}
\end{figure}
	
\begin{figure}[tb]
		\centering
		\includegraphics[width=0.48\textwidth]{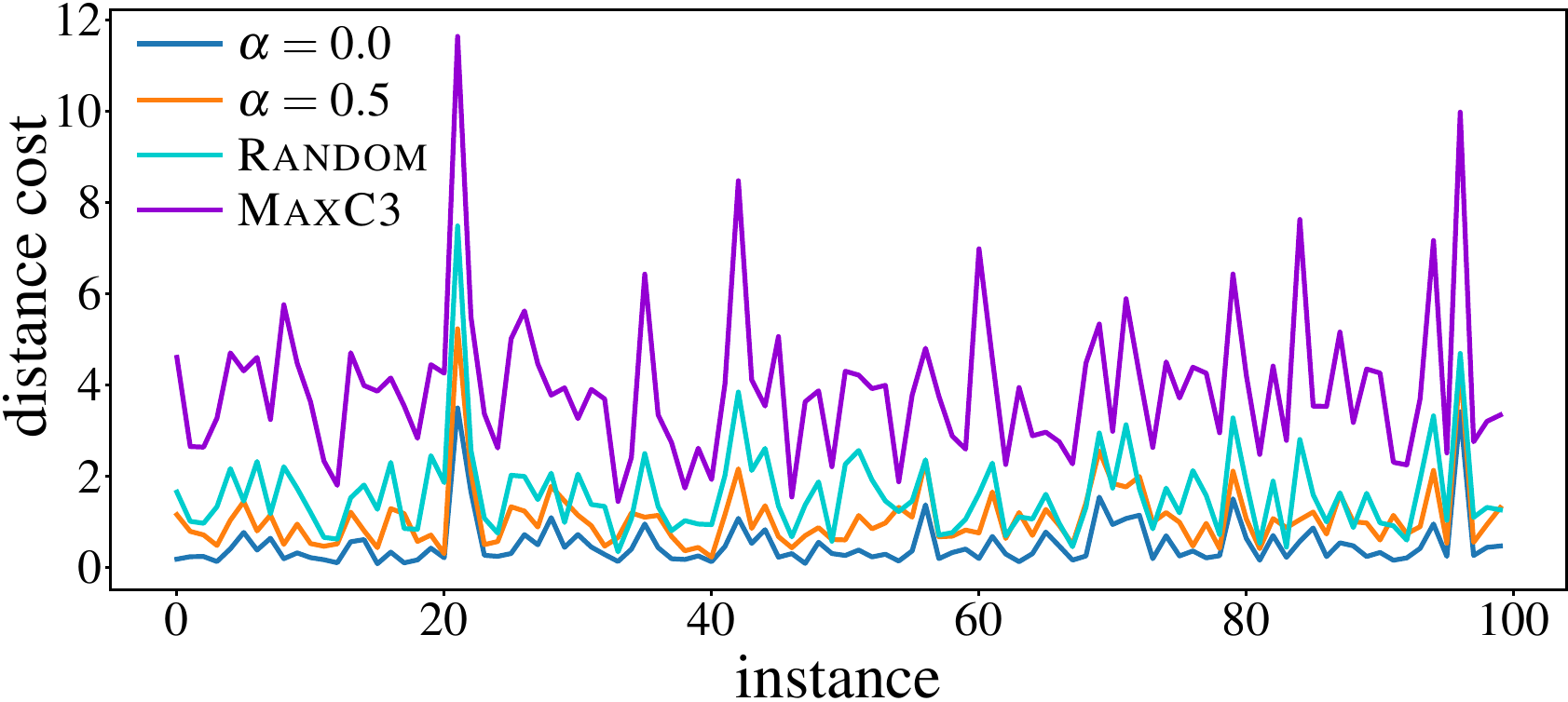}
		\caption{Optimal distance cost $c_{\mathrm{D}}$ for
                  each instance and $\alpha \in \{0.0, 0.5
                  \}$. \textsc{Random} is the distance cost of randomly generated labelings and \textsc{MaxC3} is the maximal possible distance cost.}
		\label{fig:ilp_distances}
\end{figure}

\paragraph{Evaluation of Exact Approach}

We first evaluate our exact approach with respect to both optimized
criteria, i.e., the crossing cost (C2) and the distance cost (C3). We run the approach for
$\alpha\in\{0,0.5,1\}$. For $\alpha=0$ and $\alpha=1$ we only optimize
C3 and C2, respectively. Further, we use $\alpha=0.5$ as an
intermediate value balancing both criteria. When running our exact
approach with a number of $k=5$ labels per state, the storage capacity
of the used system is exceeded. By reducing the number of labels to
$k=4$, we are able to successively run the  approach on all 100
instances. We argue that limiting the number of labels to four still
models a realistic setting, in which a high number of leader crossings
and small vertical distances may occur in the worst case.

We investigate the potential of optimization with respect to both considered
criteria. To that end, we additionally determine the
maximum possible costs $c_{\mathrm{C}}$ and $c_{\mathrm{D}}$ of both
criteria. More specifically, instead of minimizing the objective
function in our ILP formulation, we maximize it for $\alpha = 0$ and
$\alpha = 1$, respectively.  We denote the results by \textsc{MaxC2}
for $\alpha=1$ and by \textsc{MaxC3} for $\alpha=0$.
Further, we created for each instance a labeling in which features of the same weight  group are ordered randomly; we refer to them as \textsc{Random}. 
The results are found in
\autoref{fig:ilp_crossings} and \autoref{fig:ilp_distances}.
The crossing costs are displayed without normalization so that they show the absolute number of crossings.

We observe that the non-normalized leader cost $c_{\mathrm{}}$ lies between $25.8$
($\alpha=1$) and $73.5$ (\textsc{MaxC2}) on average.  Hence, by
choosing $\alpha = 1$ the number of leader crossings can be reduced by
$\delta(c_{\mathrm{C}},\mathcal S_{1},\mathcal S_{\textsc{MaxC2}})= 64.8 \%$ in maximum. When considering the trade-off 
$\alpha = 0.5$, the number of leader crossings can still be reduced by
$\delta(c_{\mathrm{C}},\mathcal S_{0.5},\mathcal S_{\textsc{MaxC2}}) =62.9 \%$ on average. Compared to a randomly generated labeling (\textsc{Random}), the number of leader crossings is reduced by $\delta(c_{\mathrm{C}},\mathcal S_{0.5},\mathcal S_{\textsc{Random}})= 44.3 \%$ on average.

Concerning the distance cost of the
optimal results for $\alpha \in \{0,0.5\}$, \textsc{MaxC3} and \textsc{Random}, we
make a similar observation; see~\autoref{fig:ilp_distances}. The
average distance cost is $4.0$ in maximum and $0.5$ in minimum, which
means that the distance cost can be reduced by $\delta(c_{\mathrm{D}},\mathcal S_{0},\mathcal S_{\textsc{MaxC3}})=87.5 \%$ in maximum. For
$\alpha = 0.5$ the distance cost is still reduced by $\delta(c_{\mathrm{D}},\mathcal S_{0.5},\mathcal S_{\textsc{MaxC3}})=72.8 \%$ on
average. Compared to a randomly generated labeling, with $\alpha = 0.5$ the distance cost is reduced by $\delta(c_{\mathrm{D}},\mathcal S_{0.5},\mathcal S_{\textsc{Random}})=33.5 \%$ on average. These results show that both considered criteria C2 and C3
can be substantially optimized by an appropriate choice of $\alpha$.

\begin{figure}[tb] \centering
\includegraphics[width=0.48\textwidth]{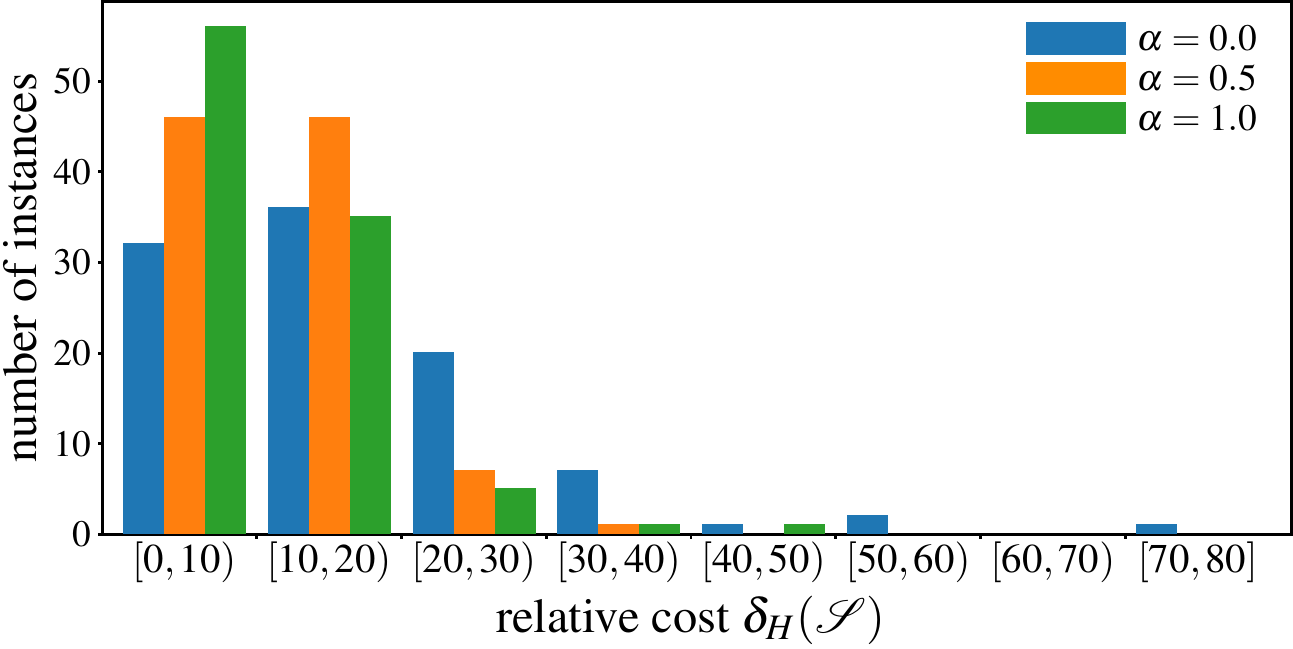}
\caption{The relative cost $\delta_H(\mathcal S)$ for the labeling
  $\mathcal S$ of the heuristic and the according optimal solution for
  $\alpha \in \{0,0.5,1\}$.}
	\label{fig:heuristicquality}
\end{figure}

\paragraph{Exact Approach vs.\ Heuristic}
We evaluate the quality of our heuristic by comparing its results with
the optimal results obtained from the exact approach. For each instance and each $\alpha \in \{0,0.5,1\}$ we
compute the relative cost $\delta_H(\mathcal S):=\delta(c_{\textrm{Slid}},\mathcal S,\mathcal S_E)$
where $\mathcal S$ is a labeling produced by the heuristic and
$\mathcal S_{E}$ is the optimal labeling for the same instance; see~\autoref{fig:heuristicquality}. For $\alpha = 0$ and for $68 \%$ of
the instances the relative cost is less than $20 \%$. Further, $92\%$
and $93\%$ of the instances have relative costs less than $20\%$ for $\alpha=0.5$ and $\alpha=1.0$, 
respectively. On average, the relative costs of the heuristic solutions are
$17.1 \%$, $11.3 \%$ and $9.5 \%$ for $\alpha=0$, $\alpha=0.5$, and $\alpha=1.0$, 
respectively. In particular, for $\alpha \in\{0.5,1.0\}$ about $90 \%$
of the quality of the optimal solution was reached on average.
Hence, apart from some instances, our heuristic provides results that are sufficiently close
to optimal.

	
	\begin{figure}[tb]
		\centering
		\includegraphics[width=0.45\textwidth]{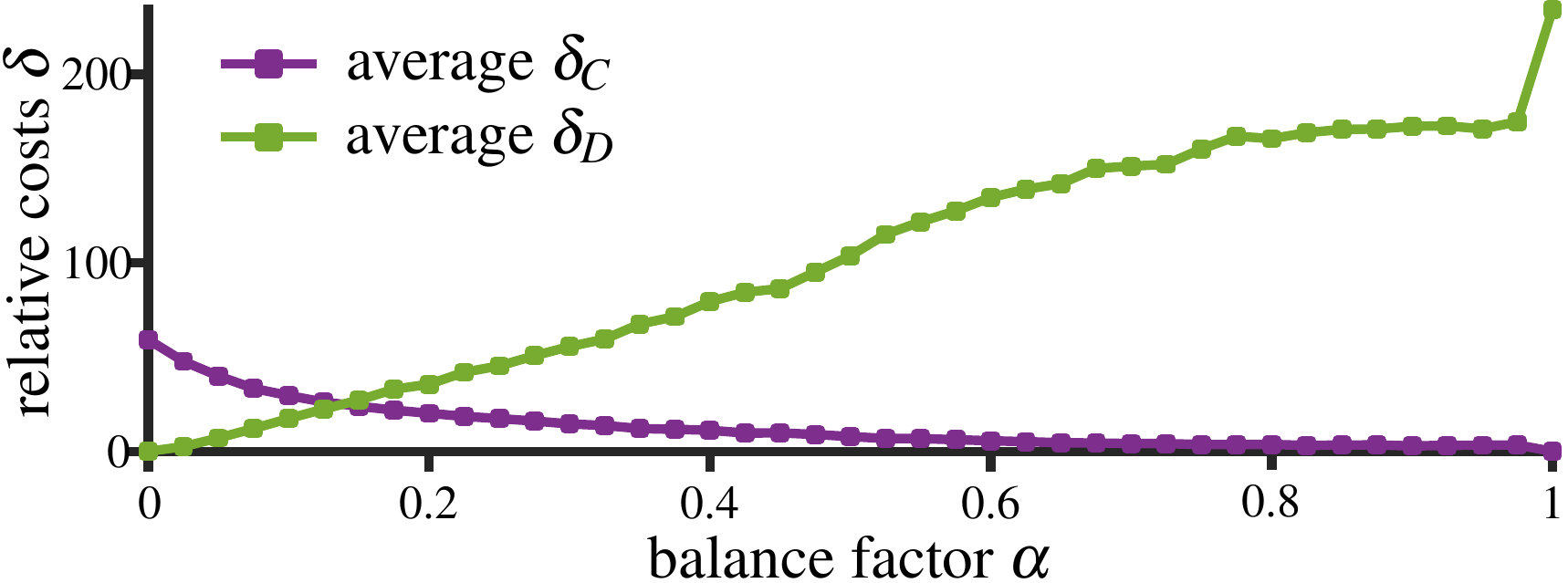}
		\caption{Average relative leader cost $\delta_{\mathrm{C}}$ and average relative feature cost $\delta_{\mathrm{D}}$ for each $\alpha \in \mathcal A$.}
		\label{fig:sliding-scoreDistances}
	\end{figure}
	
	\begin{figure}[tb] \centering
\includegraphics[width=0.48\textwidth]{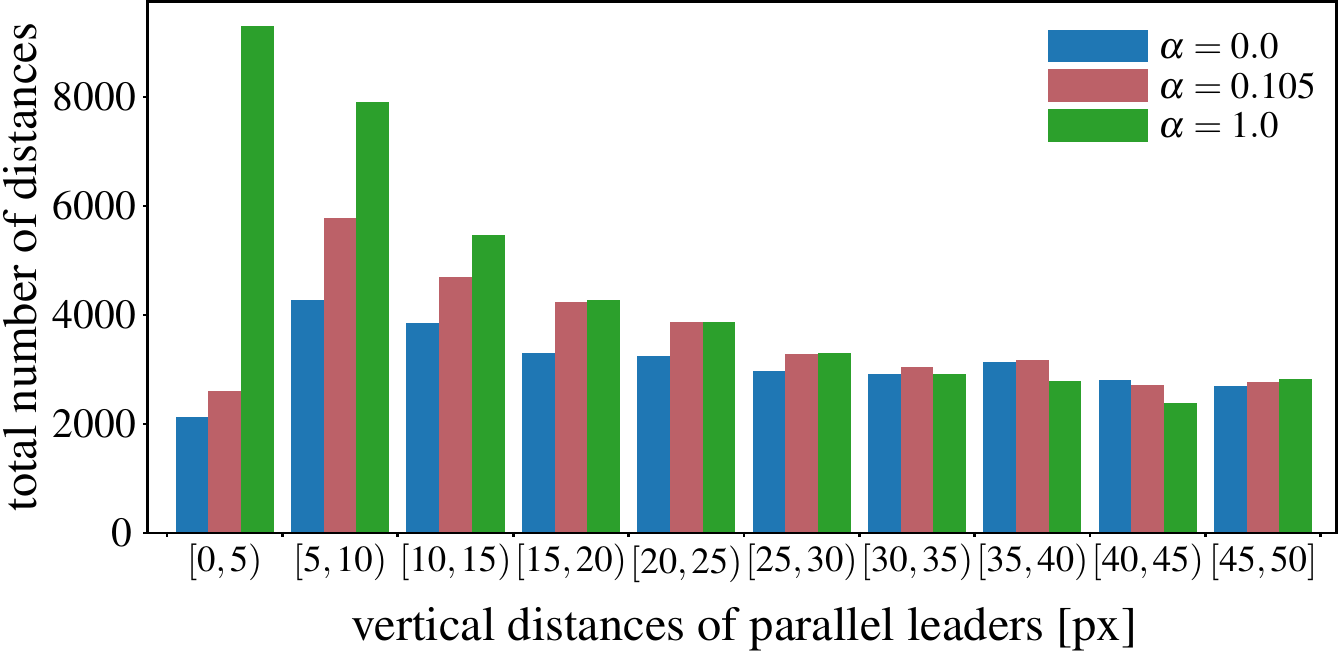}
\caption{Distribution of small vertical leader distances in the range $[0,50]$~pixels for
  $\alpha \in \{0,0.105,1\}$.}
	\label{fig:disthist}
\end{figure}

\paragraph{Heuristic}
Finally, we systematically analyze how the heuristic behaves for
each $\alpha \in\mathcal A$ (see
Section~\ref{sec:eval_mpl}). We denote the corresponding labeling by $\mathcal S_\alpha$. As we do not
take the exact approach into account, we use five labels for each
state as also done in the other labeling methods.

For evaluating Criterion C2 we consider for each labeling
$\mathcal S_\alpha$ the \emph{relative crossing cost} $\delta_{\mathrm{C}}(\mathcal S_\alpha):=\delta(c_{\mathrm{C}}\mathcal S_\alpha,\mathcal S_1)$,
where $\mathcal S_1$ only takes Criterion C2 into account; see
\autoref{fig:sliding-scoreDistances}.  As we create $\mathcal S_1$ by
means of an heuristic, the relative cost can be negative.  
Similarly, we evaluate the \emph{relative distance cost} $\delta_{\mathrm{D}}(\mathcal S_\alpha):=\delta(c_{\mathrm{D},}\mathcal S_\alpha, \mathcal S_0)$,
where $S_0$ only takes Criterion C3 into account. 
We obtain that the average relative crossing and distance costs  are almost equal for $\alpha=0.105$. We
deem this value of $\alpha$ to be a suitable compromise between 
Criteria C2 and C3. 

To further evaluate the extent to which Criterion C3 is optimized, we consider the distribution of vertical leader distances within a range of $[0,50]$ pixels for $\alpha \in \{0,0.105,1\}$; see~\autoref{fig:disthist}. We obtain that compared to the results in which only Criterion C2 is optimized ($\alpha=1$) we can reduce the number of vertical leader distances smaller than five~pixels by $77.2 \%$ when optimizing C3 ($\alpha=0$).
However, with
$\alpha = 0.105$ we can still reduce these smallest leader distances by $72.1 \%$.





\subsection{Stacking Boundary Labeling}
We analyze both the weight costs (C1)---similar as done for the
multi-page boundary labeling---and the distance cost (C3). Recall that a stacking
boundary labeling~$\mathcal S$ can be interpreted as a
multi-page boundary labeling. Hence, we can evaluate the
relative weight costs~$\delta(c_{\mathcal W} ,\mathcal S, \mathcal S_0)$ where $\mathcal S$ is a stacking boundary labeling and~$\mathcal S_0$ is the multi-page boundary labeling with minimal weight cost. On
average the relative weight cost is~$11.6 \%$ with a
standard deviation of~$8.6\%$. 
Hence, top rated features are nearly as equally distributed on the stacks
as for the optimal multi-page boundary labeling. Put differently, we obtain a 
small number of interactions for exploring them. 
Moreover, we evaluate the vertical distances of the leaders that run horizontally above each other.
Let $H$ be the set of all pairs of leaders with horizontal
segments lying above each other and for which the leaders start at different ports. The
ratio between the size of $H$ and the overall number of leader pairs
starting at different ports is $17.6\%$ on average with a standard
deviation of $17.6\%$. 
The number of leader pairs in $H$ with a vertical distance smaller 
than five pixels is $7.9\%$. Hence, on average we obtain a smaller number of leader pairs with
horizontal segments running above each other and a smaller number of leader pairs with vertical distances under five pixels than for the multi-page boundary labeling.
	
\begin{table}[tb]
  \centering
  \caption{Key properties of the three methods.  Criteria C1--C4 are enforced as \emph{hard}
    constraint or optimized as \emph{soft} constraint if reasonable.}
		\label{table:evaluation}
  \begin{tabular}{l c c c  }
    \toprule
    \textbf{Key Property}& \textbf{Multi-Page} &  \textbf{Sliding}  & \textbf{Stacking} \\
    \midrule
    C1: weight     & soft  & hard & soft  \\  
    C2: crossing   & hard  & soft & hard  \\ 
    C3: distance   & --   & soft & --  \\  
    C4: leader     & soft  & --  & soft  \\
    \midrule
    animation    & discrete & continuous & discrete \\
    change         & $k$ labels & 1 label & 1 label \\
    \bottomrule			  
  \end{tabular}
\end{table}

\subsection{Comparison}
We finally compare the labeling methods providing
decision support for deploying them in
practice; see \autoref{table:evaluation}. In all methods one criterion is a hard
constraint and two are additionally optimized. Hence, the enforced
criterion can be interpreted as the primary design rule. For example, for crossing-free labelings multi-page and stacking boundary labeling are preferable.
Further, the methods differ in their interaction techniques. Both multi-page and stacking boundary labeling are designed for discrete interaction, so that switching between two states involves no transitional animation. In contrast, sliding boundary labeling provides continuous animation, which supports the user in tracing changes.
Another difference is the amount of change between each interaction
step. For multi-page boundary labeling all $k$ labels are exchanged by
completely different labels in each interaction step.
In contrast, for
the two other labeling methods only one label is exchanged. Hence,
for multi-page boundary labeling less interaction is necessary to
explore all information. On the other hand, by means of the two other
labeling methods the user can adjust the labeling in small steps. In
sliding boundary labelings the interaction is still strongly bounded,
i.e., the labels are slid along in a specific order. Stacking
boundary labeling provides less restricted interaction. The user
can easily customize the displayed labeling by changing the topmost
labels on the stacks. 
%
\begin{table}[t]
	\centering
\renewcommand{\arraystretch}{1}
	\caption{Averaged running times with standard deviation of the heuristic approaches in milliseconds.}
		\label{table:runtimes}
\begin{tabular}{llccc}
\toprule
\multicolumn{2}{c}{\textbf{Setup}} & \multicolumn{1}{c}{\textbf{Multi-Page}} & \multicolumn{1}{c}{\textbf{Sliding}} & \multicolumn{1}{c}{\textbf{Stacking}} \\
\midrule
\multirow{2}{*}{$k=5$} & $n=30$ & 10.58 $\pm$ 2.89 & 44.30 $\pm$ 0.52 & 10.51 $\pm$ 3.51 \\\vspace{5px} 
 & $n=100$ & 83.53 $\pm$ 11.18 & 53.53 $\pm$ 0.60 & 71.10 $\pm$ 4.92 \\

\multirow{2}{*}{$k=10$}& $n=30$ & 7.67 $\pm$ 3.09 & 75.47 $\pm$ 0.93 & 8.72 $\pm$ 3.93 \\
 & $n=100$ & 85.58 $\pm$ 12.53 & 96.20 $\pm$ 1.82 & 74.06 $\pm$ 5.40 \\
\bottomrule	
\end{tabular}
\end{table}
For all methods we measured the running time of the
  heuristics for $k=5$ ports and $n=30$ point features;
  see~\autoref{table:runtimes}. For multi-page boundary labeling and sliding boundary labeling, we averaged over all choices of $\alpha$. To further assess the impact of the
  number of ports and features on the running time, we also considered
  $k=10$ and
  $n=100$. 
  We observe that for each setting the running time is less than a
  tenth of a second, which we consider reasonable in terms of actual
  applicability.
Altogether, no labeling method prevails over the others. The concrete application
decides which of them is the preferable choice. 


	
\section{Conclusion}
In this paper, we investigated three labeling methods that reduce the necessity
of zooming by providing the user with the possibility of browsing
through the labels without changing the displayed map in the
background. For all three methods we presented algorithms that run
fast enough for real-world applications on devices such as
smartwatches. They enforce and optimize the Criteria C1--C4, which we
have identified in an expert study. With a systematic and quantitative 
evaluation of the mathematical models and algorithms we laid the foundation 
for zoomless maps using external labeling. 
Altogether, this paper provides the following novel and scientific contributions.

\setdefaultleftmargin{2em}{2em}{}{}{}{}
   \begin{compactitem}
   \item New external labeling methods that allow a user to
navigate through dense sets of points of interest in zoomless maps.
   \item Design decisions for external labeling based on expert advice.
\item Generalization of the three labeling methods into a unified
     algorithmic framework.
   \item Exact algorithms and fast heuristics for the optimization problems.
  \end{compactitem}
 
So far we have considered all three
labeling methods under the assumption that the map is fixed. It is
future work to investigate how the labeling methods can be combined
with operations such as zooming and panning such that different frames
have similar labelings. In terms of actual applicability, future work should additionally focus on how well the methods can be translated to interaction on an actual small-screen device.
We finally emphasize that the presented
methods are only a starting point towards more sophisticated labeling
concepts. 

%

	\acknowledgments{
		Partially funded by the Deutsche Forschungsgemeinschaft (DFG, German Research Foundation) under Germany’s Excellence Strategy – EXC 2070 – 390732324.
		Partially funded by Zoomless Maps: Models and Algorithms for the Exploration of Dense Maps with a fixed Scale of the German Research Foundation (DFG) [grant number 5451/6-1]}
	
	\bibliographystyle{abbrv}

	\bibliography{strings,references}
\end{document}